\documentclass[12pt,preprint]{aastex}
\usepackage{amssymb}

\def\h0 {$h_0$=71 km s$^{-1}$ Mpc$^{-1}$}
\def\ergs { erg s$^{-1}$}
\newcommand{\be}{\begin{equation}}
\newcommand{\ee}{\end{equation}}
\newcommand{\ce}{\ifmmode {\cal E} \else ${\cal E}$\ \fi}
\newcommand{\kms}{\ifmmode {\rm km\ s}^{-1} \else km s$^{-1}$\ \fi}
\newcommand{\tes}{\ifmmode \tau_{\rm es} \else $\tau_{\rm es}$\ \fi}
\newcommand{\tk}{\ifmmode \tau_{\rm K} \else $\tau_{\rm K}$\ \fi}
\newcommand{\vfwhm}{\ifmmode V_{\mbox{\tiny FWHM}} \else
            $V_{\mbox{\tiny FWHM}}$\fi}
\newcommand{\msun}{\ifmmode M_{\odot} \else $M_{\odot}$\ \fi}
\newcommand{\afe}{\ifmmode {\mathcal A_{\rm Fe}} \else${\mathcal A_{\rm Fe}}$\ \fi}

\newcommand{\lb}{\ifmmode L_{\rm Bol} \else $L_{\rm Bol}$\ \fi}
\newcommand{\ledd}{\ifmmode L_{\rm Edd} \else $L_{\rm Edd}$\ \fi}
\newcommand{\lx}{\ifmmode L_{\rm 2-10keV} \else  $L_{\rm 2-10keV}$\ \fi}
\newcommand{\hb}{\ifmmode H\beta \else H$\beta$\ \fi}
\newcommand{\mbh}{\ifmmode M_{\rm BH}  \else $M_{\rm BH}$\ \fi}
\newcommand{\lv}{\ifmmode \lambda L_{\lambda}(5100\AA) \else $\lambda L_{\lambda}(5100\AA)$\ \fi}

\def\astrobj#1{#1}

\newcommand{\heii}{He {\sc ii}\ }

\def\ariel5{{\it Ariel 5}\ }

\def\heao1{{\it HEAO~1}\ }

\shorttitle{Black hole masses of ULXs} \shortauthors{Xin-Lin Zhou}

\begin{document}
\title{On the Black Hole Masses In Ultra-luminous X-ray Sources}

\author{
Xin-Lin Zhou$^{1}$}

\affil{
$^1$ National Astronomical
Observatories, Chinese Academy of Sciences, 20A Datun Road, Chaoyang District, 100012, Beijing, China \\
}
\email{zhouxl@nao.cas.cn}


\begin{abstract}
Ultra-luminous X-ray sources (ULXs) are off-nuclear X-ray sources in
nearby galaxies with X-ray luminosities $\geq$ 10$^{39}$ erg
s$^{-1}$. The measurement of the black hole (BH) masses of ULXs is a
long-standing problem. Here we estimate BH masses in a sample of
ULXs with XMM-Newton observations using two different mass
indicators, the X-ray photon index and X-ray variability amplitude
based on the correlations established for active galactic nuclei
(AGNs). The BH masses estimated from the two methods are compared
and discussed. We find that some extreme high-luminosity ($L_{\rm X}
>5\times10^{40}$ erg s$^{-1}$) ULXs contain the BH of 10$^{4}$-10$^{5}$ $M_\odot$.
The results from X-ray variability amplitude are in conflict with
those from X-ray photon indices for ULXs with lower luminosities.
 This suggests that these ULXs generally accrete at rates different
 from those of X-ray luminous AGNs, or they have different power spectral
 densities of X-ray variability. We conclude that most of ULXs accrete at
super-Eddington rate, thus harbor stellar-mass BH.
\end{abstract}


\section{Introduction}
A population of compact X-ray sources in nearby galaxies  with
observed isotropic X-ray luminosities  $\geq$ 10$^{39}$ erg s$^{-1}$
are called Ultra-luminous X-ray sources (ULXs; Makishima et al.
2000). It is widely accepted that ULXs are accreting black holes
(BHs; Fabbiano 2006; Roberts 2007). ULXs may contain
 3 classes of BHs:  normal stellar mass BHs ($\sim$10 $M_\odot$),
 massive stellar BHs ($\le$ 100 $M_\odot$ ), and intermediate
 mass BHs (IMBH; $10^2-10^4$ $M_\odot$; Feng \& Soria 2011).

There is a lack of reliable method to measure the BH masses for
ULXs. Long-term observations of optical counterparts of ULXs might
provide a dynamical BH mass measurement. The broad \heii 4686 line
may change its velocity between two observations (Pakull et al.
2006). There is only one dynamical measurement for the BH mass for
\astrobj{M 101 ULX-1} (Liu et al. 2013). Generally, the optical
lines were found to vary randomly (Roberts et al. 2011). The
accretion disk temperature derived from X-ray spectral fitting will
place constraints on BH masses for ULXs (Miller et al. 2003).
However, the real disk temperature is controversial because the disk
blackbody fitting may be unphysical (Stobbart et al. 2006). Radio
observations could constrain BH mass via the fundamental plane
(Merloni et al. 2003; Falcke et al. 2004). However, radio
observations are generally difficult in resolving the point source
of a ULX and the error of mass estimates is large (Webb et al. 2012;
Cseh et al. 2012).

There are well-established correlations between X-ray photon index
and Eddington ratio, X-ray Variability Amplitude (XVA; also known as
``excess variance'';
 Nandra et al. 1997) and BH mass, in X-ray luminous active galactic nuclei (AGNs). We
 apply these correlations to estimate BH masses for ULXs and check
 consistency of the results. We find that ULXs generally accrete at different rates
 from X-ray luminous AGNs except some extreme high-luminosity ULXs
 ($L_{\rm X}> 5\times$ $10^{40}$ erg s$^{-1}$). Throughout this work, we assume the
concordance cosmology of \h0 , $\Omega_{m}=0.27$, and
$\Omega_{\Lambda}=0.73$.

\section{$M_{\rm BH}$ estimated from $\Gamma$ and $L_{\rm 2-10keV}$}
X-ray spectral studies of accreting BHs including Galactic X-ray
binaries (XRBs) and AGNs show complex correlations between the
photon index, $\Gamma$ of the power-law component (generally
measured over the $2-10$ keV range, $A(E)\propto E^{-\Gamma}$) and
the accretion rate in unit of the Eddington ratio, $L_{\rm
bol}/L_{\rm Edd}$, where $L_{\rm bol}$ is the bolometric luminosity
and $L_{\rm Edd}$ is the Eddington luminosity, over a large range
(Lu \& Yu 1999; Wu \& Gu 2008; see Figure 1). At the very low
$L_{\rm bol}/L_{\rm Edd}$, $\Gamma$ flattens as $L_{\rm bol}/L_{\rm
Edd}$ increases (Corbel et al. 2008; Gu \& Cao 2009).
 Above a transition point in the range of $10^{-2}-10^{-3}$ (Younes et al. 2011; Qiao \&
 Liu 2013), $\Gamma$ increases as $L_{\rm bol}/L_{\rm Edd}$ increases (Wang et al. 2004; Shemmer et al. 2006; Zhou \&
Zhao 2010; Jin et al. 2012).  At the super Eddington region ($L_{\rm
bol}/L_{\rm Edd}>1$), we do not
 know how the  $\Gamma$ behaves but it might turn systematically flatter than the
  correlations predict (Ai et al. 2011; Kamizasa et al. 2012).

Zhou \& Zhao (2010) present a correlation between $\Gamma$ and both
Eddington ratio  and bolometric correction,
 based on simultaneous X-ray, UV, and optical observations of reverberation-mapped AGNs
 as follows,
\begin{equation}
\log\left(L_{\rm bol}/L_{\rm Edd}\right) = (2.09 \pm 0.58)\,\Gamma
- (4.98 \pm 1.04);
\end{equation}\\[-30pt]
\begin{equation}
\log\left(L_{\rm bol}/L_{\rm 2-10keV}\right) = (1.12 \pm
0.30)\,\Gamma - (0.63 \pm 0.53).
\end{equation}

We assume that ULXs follow the same correlations. If ULXs accrete at
the rate of $0.004< L_{\rm bol}/L_{\rm Edd}<1$ (Part II in Figure
1), we can estimate $M_{\rm BH}$ ($L_{\rm Edd}$)  from $\Gamma$ and
$L_{\rm 2-10keV}$ using Equations (1) and (2).

Sutton et al. (2012) presented a sample of eight extreme luminosity
ULXs (with $L_{\rm X} > 5\times 10^{40}$ erg s$^{-1}$) selected from
the cross-correlation of the 2XMM-DR1 and RC3 catalogues. These
objects represent the high-luminosity end of ULX populations. We
estimate $M_{\rm BH}$
 for the ULX sample with extreme luminosities (Table 1).
The results show that log($M_{\rm BH}/M_\odot$) is in the range of $4.5-5.8$ for these objects.

 Gonz\'alez-Mart\'in et al. (2011) collected all bright ULXs from the
literatures (Heil et al. 2009; Galdstone et al. 2009). This sample
consists of 15 ULXs with currently available best-quality XMM-Newton
data. We also estimate $M_{\rm BH}$ for these objects. We find that
log($M_{\rm BH}/M_\odot$) is in the range of $2.8-4.5$ (Table 2).

\section{$M_{\rm BH}$ estimated from XVA }

The X-ray Variability Amplitude (XVA), $\sigma^2_{\rm rms}$, is the
 variance of a light curve normalized by its mean squared after correcting
 for experimental noise (Nandra et al. 1997),
\begin{equation}
\sigma^2_{\rm rms} = \frac{1}{N\mu^2}\,\sum_{i=1}^{N}
\left[\left(X_i - \mu\right)^2 - \sigma_i^2\right].
\end{equation}

Zhou et al. (2010) present the anti-correlation
 between $M_{\rm BH}$ and XVA in the 2-10 keV band derived from a sample of
reverberation mapping AGNs,
  \begin{equation}
M_{\rm BH} = 10^{4.97\pm0.26}\, \left(\sigma^2_{\rm rms}\right)^{-1.00\pm0.10}\, M_{\odot}.
\end{equation}

It was found that this anti-correlation shows a small intrinsic
dispersion, no larger than the uncertainties in $M_{\rm BH}$ (Zhou
et al. 2010; Ponti et al. 2012; Kelly et al. 2013). This correlation
can be explained by a universal broken power-law shape of the power
spectral density (PSD) function from XRBs to AGNs (Markowitz et al.
2003; McHardy et al. 2004; Zhou et al. 2010).

 Gonz\'alez-Mart\'in
et al. (2011) calculated the XVA in 2-10 keV band  for the sample
with best X-ray data quality (listed in Table 2). It was found that
the XVAs for this sample are generally quite small. Assuming these
objects have a universal shape of the PSD (and thus Equation (4) is
applicable), $M_{\rm BH}$ can be estimated from the XVA. The
resulting value of $M_{\rm BH}$ are quite large, comparable to those
of AGNs. If this is true, some ULXs might accrete at a very low
rate. But this result is in conflict with that from X-ray photon
indices, which predict a log($M_{\rm BH}/M_\odot$) for these objects
in the range of $2.8-4.5$ (See Table 2).

\section{Discussion}
The BH masses estimated from the X-ray photon indices for
extreme-luminosity
 ULXs (Table 1) are in range of $10^{4}-10^{5}$ $M_\odot$. This is in agreement with the
 results  from their radio observations (Mezcua et al. 2013).
Thus these objects are very likely to harbor massive BH, with the
same accretion rate as luminous AGNs.

The XVA are small for objects with the best X-ray data quality
(Table 2), this leads to very large $M_{\rm BH}$ estimates using
XVA. It was found that extreme-luminosity ULXs (objects in Table 1)
show larger fractional X-ray variability (square root of XVA) than
low-luminosity ULXs in the $0.3-10$ keV band (Sutton et al. 2012).
It is very unlikely that the extreme-luminosity ULXs generally have
smaller $M_{\rm BH}$ than low-luminosity ULXs. It is very likely
from these data that there are different populations among ULXs (and
thus different PSD shape). However, the PSD shape of ULXs could not
be constraint well with the current data (Heil et al. 2009). It was
suggested that some ULXs in Table 2 may have a PSD with a doubly
power-law shape (Gonz\'alez-Mart\'in et al. 2011).

It has been suggested that the observed characteristics of the
source depend on both accretion rate and the inclination of the ULX
system (Sutton et al. 2013). The accretion rate of a source should
be the key factor. The simplest conclusion we can draw from those
spectral and timing results is that it is not in the  $0.004< L_{\rm
bol}/L_{\rm Edd}<1$ regime where both correlations apply for objects
in Table 2. There is an additional reason why the data suggest we
are not in the $0.004< L_{\rm bol}/L_{\rm Edd}<1$ regime: very few
or none of those ULXs is in the high-soft state, as we would expect
at those values of $L_{\rm bol}/L_{\rm Edd}$.

The $\Gamma$ and XVA depend on the ratio between power in the
Compton cloud and power in the accretion disk. For $L_{\rm
bol}/L_{\rm Edd}$ below the transit
 point, the hot inner flow or  the jet becomes slightly more dominant as $L_{\rm bol}/L_{\rm Edd}$
 increases (Gardner \& Done 2013), so  $\Gamma$  goes down up to $L_{\rm bol}/L_{\rm Edd}$  $\sim$ 0.004.
 Then a cold disk becomes more important, so $\Gamma$  increases for
 $0.004 <L_{\rm bol}/L_{\rm Edd}<1$ (Qiao \& Liu 2013; Cao \& Wang 2014). Then the scattering region (e.g.
 outflows in slim disk models) becomes more dominant again, so it is expected $\Gamma$ to
 decrease again for $L_{\rm bol}/L_{\rm Edd} >1$ (See Figure 1).

$\Gamma$ in some ULXs decreases as the X-ray flux increases, for an
instance, \astrobj{NGC 1313 X-2} (Feng \& Kaaret 2006; Wu \& Gu
2008). This suggests that ULXs are either at $L_{\rm bol}/L_{\rm
Edd} <~0.004$, or $>~1$ (see Figure 1). A low XVA is also associated
with a dominant scattering region/outflows, so ULXs with low XVA are
probably consistent with either  $L_{\rm bol}/L_{\rm Edd}<~0.004$,
or $>~1$. It is hard to distinguish the spectral and timing
properties from the hard state $L_{\rm bol}/L_{\rm Edd}<~0.004$, or
the super-Eddington state $>~ 1$. They are both dominated by
inverse-Compton scattering over thermal disk emission. They both
have outflows and low variability.

A few objects may favor the interpretation that ULXs are at $L_{\rm
bol}/L_{\rm Edd} <~ 0.004$.
 We are in favor of the other scenario, $L_{\rm bol}/L_{\rm Edd} >~ 1$, for most of objects in Table 2.
 we exclude the $0.004 <~ L_{\rm bol}/L_{\rm Edd} <1$ regime for most of objects.
so we generally cannot directly apply those correlations to estimate
BH mass. There are  6 reasons why we  think most of ULXs are more
likely to be at $L_{\rm bol}/L_{\rm Edd} >~ 1$ (that is, $M_{\rm
BH}$ $<~ 100 ~M_\odot $) rather than $L_{\rm bol}/L_{\rm Edd}  <~
0.004$ (that is, $M_{\rm BH}$ $>~ 10000 ~M_\odot $) as follows:

a) these ULXs in star-forming galaxies follow the same luminosity
function as (stellar) high-mass X-ray binaries (Swartz et al. 2011)
. ULXs in old galaxies follow the same luminosity distribution as
low-mass X-ray binaries (Kim \& Fabbiano 2004; Swartz et al. 2011).
Their number and luminosity matched exactly (see the review by Feng
\& Soria 2011).  It would therefore be an incredible coincidence if
X-ray binaries and ULXs were completely different objects.

b) ULXs have the same spatial distribution (number of sources at a
given distance from the centre of a galaxy) as X-ray binaries
(Swartz et al. 2011). Again, it would be an amazing coincidence if
X-ray binaries (formed in situ) and BHs accreted from outside had
the same radial distribution.

c) ULXs are more frequently found in starburst galaxies, associated
with a high star formation rate (Fabbiano 2006). This suggests that
they are formed via massive star formation. There is no reason why
accretion of satellite dwarf galaxies should favor starburst
systems.

d) In terms of specific frequency (number per unit galaxy mass),
ULXs are more frequent in small disk galaxies ($M* \sim 1E10$
$M_\odot$, spectral type Sc) rather than larger spirals or
ellipticals (Swartz et al. 2011) . However, if ULXs came from dwarf
accretion, they should be more abundant in those types. Scd galaxies
do not have much satellite accretion.

e) Figure 2 shows the X-ray luminosities against the optical V-band
luminosities for some ULXs and LINERs. ULXs (open circle) generally
have a large ratio of $L_{\rm X}/L_{\rm opt}$ of about 100, like in
stellar BHs with a donor star (Tao et al. 2011). Low-luminosity AGNs
(filled circle) may have the same X-ray luminosity ($L_{\rm X} \sim$
a few$\times 10^{40}$ erg s$^{-1}$) but they tend to have
 more luminous optical nuclei, $L_{\rm X}/L_{\rm opt} \sim 0.1-10$ (Younes et al. 2012).

f) If many ULXs had $M_{\rm BH} >10^4~M_\odot$ , and came from
accreted dwarfs, we would expect a sign of a galaxy merger or
disruption around them. A BH with $M_{\rm BH} >10^4~M_\odot$ lives
in a massive globular cluster, or perhaps in an Scd galaxy with a
nuclear star cluster (Mapelli et al. 2012). But most ULXs do not
show any evidence of a disrupted galaxy around them, and they are
generally not located in massive star clusters. This does not rule
out the merger scenario but does not give any evidence in favor of
it.

It is possible that some ULXs might be off-nuclei analogs of
low-luminosity AGNs. Most likely, some ULXs behave like LINER
sometimes, when they have the same luminosity and ionize the
surrounding medium. Some ULXs are found to be associated with weak
H$\alpha$ line (Wiersema et al.
 2010; Cseh et al. 2012). Pakull et al. (2010) reported an
 off-nucleus
super radio bubble in the galaxy \astrobj{NGC 7793}.
 Generally, most of ULXs lack such
observations. If future large optical and radio telescopes, such as
TMT, ALMA, SKA could reveal low-ionization emission lines associated
with some ULXs, we call such objects Low Ionization Off-Nuclear
Emission line Region (LIONER).

\section{Conclusions}
Under the assumption that ULXs are accreting at the same rates as
 X-ray luminous AGNs,  we estimate black hole masses for ULXs
 using the X-ray photon index and X-ray variability amplitude and check consistence
of the results. We find the results from X-ray variability amplitude
are in conflict with these from X-ray photon indices, except for the
objects with extreme luminosities.  This suggests
 that ULXs  generally accrete at rates different from those of
 X-ray luminous AGNs, or they have different power spectral
 densities of X-ray variability. We conclude that most of ULXs accrete at super-Eddington rate,
 thus harbor steller-mass
black holes. The objects with extreme luminosities may be massive
black holes ($10^{4}-10^{5}$ $M_\odot$), with the same range of
accretion rates as luminous AGNs.

\acknowledgments We thank the referee for English corrections. This
research has made use of observations obtained with XMM-Newton, an
ESA science mission with instruments and contributions directly
funded by ESA member states and the National Aeronautics and Space
Administration (NASA). This research has also made use of the NED
which is operated by the Jet Propulsion Laboratory, California
Institute of Technology, under contract with NASA. We thank Yuan, W.
and Soria, R. for reading the manuscript and give many helpful
comments to improve the manuscript significantly.
 We also thank useful discussions with Qiao, E. L.,  Liu, B. F., Xu, D. W., Liu, Z.,
Yao, S., Pan, H. W., Lu, Y. J., Gou, L. J, Ferland, G. J. and Ward,
M. This work was supported by the National Natural Science
Foundation of China under grant No. 11003022 and the XTP project
XDA04060604.

\clearpage

\begin{figure}
\includegraphics[width=11 cm, angle=270]{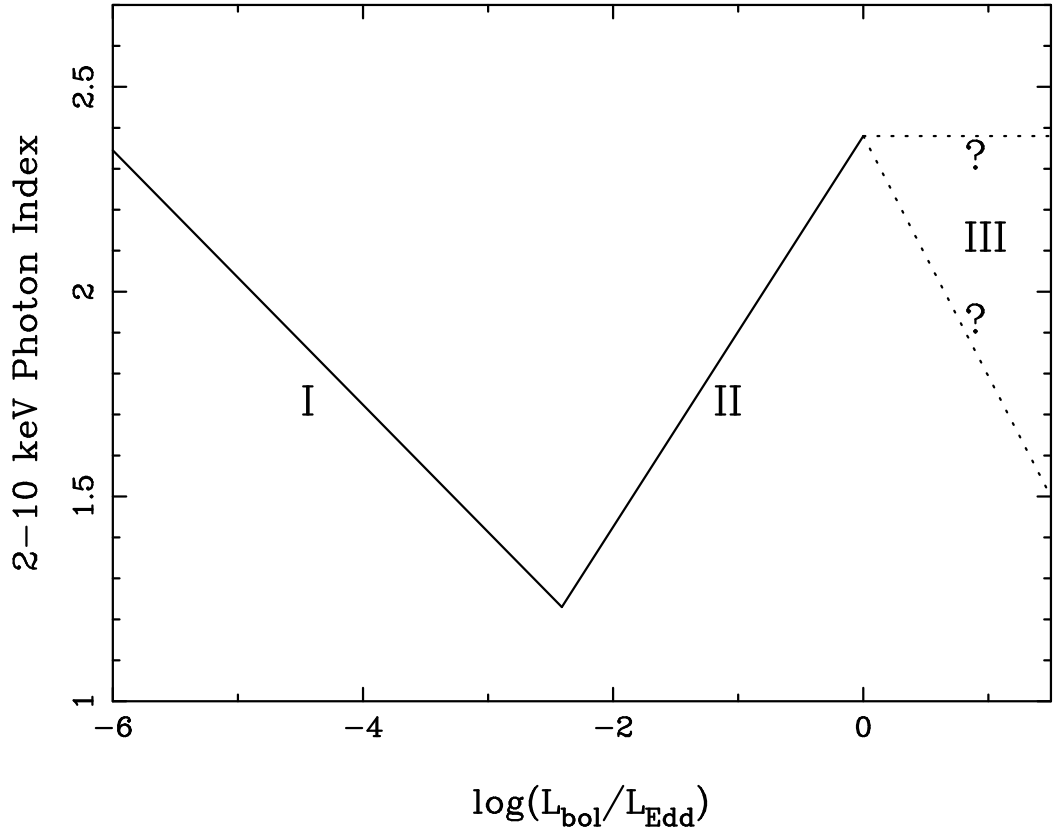}\label{fig1}
\caption  {Complex correlations between X-ray photon index and
Eddington ratio over a large range. Part I denotes the correlation
for low-luminoisty AGNs (Younes et al. 2011); Part II denotes the
correlation derived from reverberation-mapping AGNs (Zhou \& Zhao
2010); The transit point is at $L_{\rm bol}/L_{\rm Edd}\sim0.004$.
Part III denotes the correlation for super-Eddington sources.}
\end{figure}

\begin{figure}
\includegraphics[width=11 cm, angle=270]{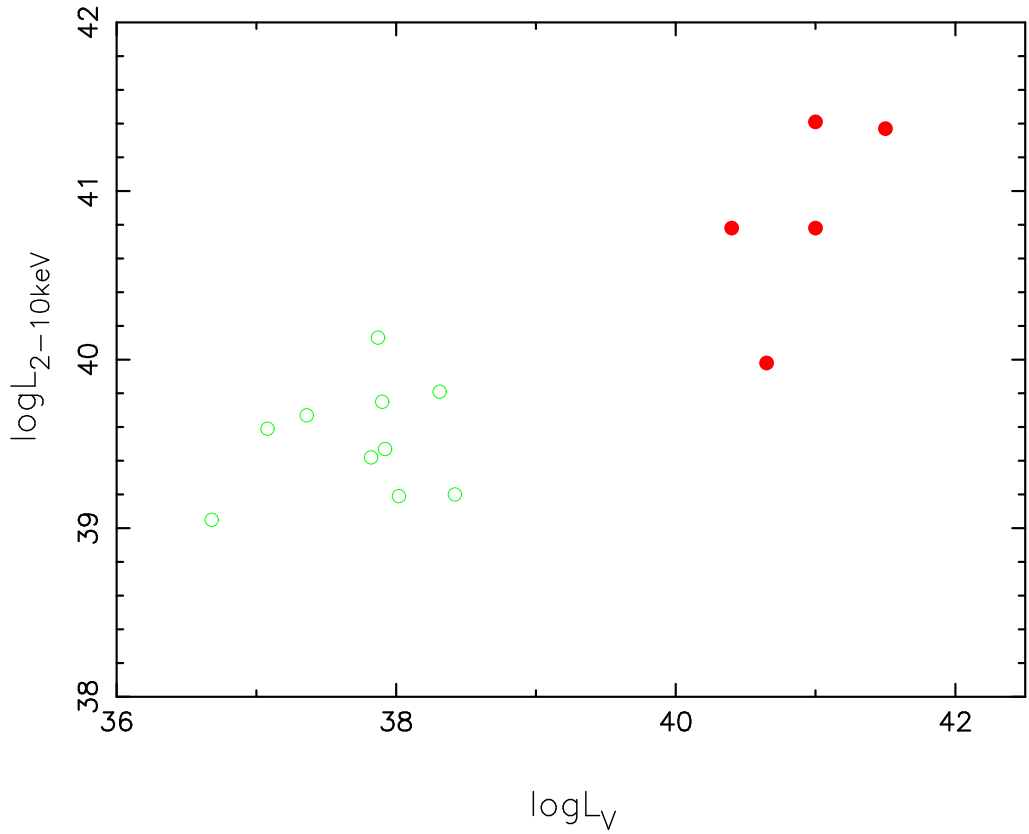}\label{fig2}
\caption  {The X-ray luminoisties against the optical V-band
luminosities for some ULXs and LINERs. ULXs (open circle) generally
have large ratio of $L_{\rm X}/L_{\rm opt}$ of about 100, like in
stellar
 BHs with a donor star (Tao et al. 2011). LINERs (filled circle) tend to have
 more luminous optical nuclei, $L_{\rm X}/L_{\rm opt} \sim 0.1-10$ (Younes et al. 2012).}
\end{figure}


\begin{table*}
\caption{Sutton et al. (2012)'s ULX sample with extreme luminosities}
\centering
\begin{tabular}{lccccccr}
\hline 2XMM source  & D(Mpc) & ${L_{\rm X,
max}}^a$ & $L_{\rm 2-10keV}^b$   & $\Gamma$   &log$M_{\rm
BH}^{\Gamma}$
\\
\hline
 2XMM J011942.7+032421    & 32.7 & $15.3 \pm 0.8$ &$7.6\pm0.7$
             &$2.2\pm0.2$      & $5.00\pm0.3$
\\
2XMM J024025.6−082428    & 18.9 & $5.1 \pm 0.2$& $2.6\pm0.2$
             &$2.2\pm0.2$   &  $4.53\pm0.23$
\\
2XMM J072647.9+854550    & 33.3 & $6.4 \pm 0.3$ & $3.5\pm0.4$ &
$1.9\pm0.3$&$4.95\pm0.50$
\\
2XMM J121856.1+142419   & 33.2 & $6.0 \pm 0.3$   & $3.8\pm0.3$&
             $1.5\pm0.1$ & $5.37\pm0.2$
\\
 2XMM J134404.1−271410   & 95.1 & $28.2 \pm 2.4$      & $15.1\pm2.5$&
$1.7\pm0.3$&$5.78\pm0.35$
\\
2XMM J151558.6+561810   & 14.9 & $4.19^{+0.09}_{-0.1}$  &
             $2.2\pm0.12$& $1.88^{+0.07}_{-0.06}$ & $4.77\pm0.20$
\\
2XMM J163614.0+661410  & 96.2 & $ 6.5 \pm 1.2$  &$3.5\pm1.0$ &
             $1.8^{+0.4}_{-0.3}$ & $5.05\pm0.40$
\\
2XMM J230457.6+122028  & 32.8 & $7.1 \pm 0.3$  &$3.4\pm0.3$ &
             $1.9^{+0.2}_{-0.1}$ & $4.94\pm0.31$
\\
\hline
\end{tabular}
\begin{minipage}{\linewidth}
Notes: $^a$Maximum detected 0.2-12 keV luminosity of the ULX
candidate, based on the 2XMM broad band fluxes and the quoted
distance, in units of $10^{40}$ \ergs. $^b$ 2-10 keV luminosity of
the ULX candidate, in units of $10^{40}$ \ergs. $M_{\rm BH}$ is
estimated from $\Gamma$  and $L_{\rm 2-10keV}$ based on the
correlations derived from radio-quiet reverberation mapping AGNs
(Zhou \& Zhao 2010).
\end{minipage}
\label{GULX}
\end{table*}

\begin{table*}
\caption{Gonz\'alez-Mart\'in et al. (2011)'s ULX Sample with
best-quality data} \vglue 0.1cm \label{tab:XVA}
\centering
\begin{tabular}{l r c c c c c}     
\hline
Source Name  & $\rm{\sigma^{2}_{rms}}$   & Data Segment &log$L_{\rm 2-10 keV}$  &  log$M_{\rm BH}^{\rm XVA}$ &log$M_{\rm BH}^{\Gamma}$   \\
          &       ($\times 10^{-3}$)      &                       \\  \hline
    NGC55ULX& $\rm{147\pm 2.1}$ & 1 & 38.7   & $5.80\pm0.15$   & $2.83\pm0.50$  \\
 NGC253PSX-2& $\rm{6.5\pm 1.3}   $   & 4 &  39.4  & $7.15\pm1.12$   & $3.53\pm0.45$   \\
  NGC1313X-1& $\rm{2.0\pm 0.9}$   & 4 &  39.6    & $7.66\pm2.8$     & $ 3.73\pm0.35 $    \\
  NGC1313X-2& $\rm{16.2\pm 0.9}$   & 4 & 39.5     & $6.76\pm0.35$   & $3.63\pm0.55$  \\
  NGC2403X-1 & $\rm{<8.3}$  & 3 &  39.2  & ...    & $3.33\pm0.5$     \\
     HoIIX-1 & $\rm{  1.3\pm 1.0}$   & 1  &  39.6 & $7.85\pm5.0$ & $3.30\pm0.36$   \\
      M81X-6 & $\rm{  4.9\pm 1.5}$   & 3  &  39.5  & $7.27\pm2.6$ & $3.41\pm0.67$    \\
      M82X-1 & $\rm{  0.9\pm 0.2}$   & 3 &  40.4  & $8.02\pm2.0$  & $4.11\pm0.65$  \\
     HoIXX-1& $\rm{  0.6\pm 0.4}$ & 3 &  39.8   & $8.19\pm5.6$    & $4.06\pm0.46$  \\
  NGC3628X-1 & $0.6\pm6.2$  & 1 & 40.0            & ...        & $3.80\pm0.53 $  \\
  NGC4559X-1& $\rm{13.0\pm 6.6}$ & 1 &  39.6     & $6.86\pm4.6$   & $3.52\pm0.36$  \\
  NGC4945X-2& $\rm{<27.0}$   &1 &  39.0      & ...    & $3.10\pm0.51$     \\
  NGC5204X-1& $\rm{1.6\pm 2.5}$   & 1 & 39.6   & ... & $3.68\pm0.35$   \\
  NGC5408X-1&  $\rm{10.2\pm 3.0}$  & 4 & 39.4   & $6.96\pm2.2$  & $3.60\pm0.45$ \\
       POX52& $\rm{93\pm 11}$  & 3 &  40.7    & $6.00\pm1.1$ & $4.55\pm0.58$  \\

\hline
\end{tabular}
\begin{minipage}{\linewidth}
Notes: Assuming a universal broken power-law shape of PSD, $M_{\rm
BH}^{\rm XVA}$ is estimated from $\rm{\sigma^{2}_{rms}}$ based on
the correlation derived from reverberation mapping AGNs (Zhou et al.
2010). $M_{\rm BH}^{\Gamma}$  is the BH mass estimated from $L_{\rm
2-10 keV}$ and $\Gamma$ using the correlations from radio-quiet
reverberation mapping AGNs (Zhou \& Zhao 2010).
\end{minipage}
\end{table*}

\end{document}